\documentclass[showpacs,amsmath,amssymb,aps,showkeys,floatfix,prd,a4paper]{revtex4}

\usepackage{dcolumn}
\usepackage{bm}
\usepackage{epsfig}
\usepackage{amsfonts}
\usepackage{amssymb,amscd}

\def\lsim{\raise0.3ex\hbox{$<$\kern-0.75em\raise-1.1ex\hbox{$\sim$}}}

\def\gsim{\raise0.3ex\hbox{$>$\kern-0.75em\raise-1.1ex\hbox{$\sim$}}}

\newcommand{\be}{\begin{equation}}

\newcommand{\ee}{\end{equation}}

\def\beq{\begin{equation}}

\def\eeq{\end{equation}}

\def\beqa{\begin{eqnarray}}

\def\eeqa{\end{eqnarray}}

\newcommand{\ba}{\begin{eqnarray}}

\newcommand{\ea}{\end{eqnarray}}

\def\gappeq{\mathrel{\rlap {\raise.5ex\hbox{$>$}}

{\lower.5ex\hbox{$\sim$}}}}

\def\lappeq{\mathrel{\rlap{\raise.5ex\hbox{$<$}}

{\lower.5ex\hbox{$\sim$}}}}

\def\Toprel#1\over#2{\mathrel{\mathop{#2}\limits^{#1}}}

\begin{document}

\title{Tau polarization effects in $\nu_\tau / \bar{\nu}_\tau $ - tungsten interactions at the LHC energies}

\author{Reinaldo {\sc Francener}}
\email{reinaldofrancener@gmail.com}
\affiliation{Instituto de Física Gleb Wataghin - Universidade Estadual de Campinas (UNICAMP), \\ 13083-859, Campinas, SP, Brazil. }

\author{Victor P. {\sc Gon\c{c}alves}}
\email{barros@ufpel.edu.br}
\affiliation{Institute of Physics and Mathematics, Federal University of Pelotas, \\
  Postal Code 354,  96010-900, Pelotas, RS, Brazil}

\author{Diego R. {\sc Gratieri}}
\email{drgratieri@id.uff.br}
\affiliation{Escola de Engenharia Industrial Metal\'urgica de Volta Redonda,
Universidade Federal Fluminense (UFF),\\
 CEP 27255-125, Volta Redonda, RJ, Brazil}
\affiliation{Instituto de Física Gleb Wataghin - Universidade Estadual de Campinas (UNICAMP), \\ 13083-859, Campinas, SP, Brazil. }

\begin{abstract}
Recent studies have demonstrated that the tau produced in charged current neutrino deeply inelastic interactions at the GeV -- TeV neutrino energy range is not fully polarized. In this paper we investigate the impact of the tau polarization on the pions generated in its decay. In particular, we consider tau neutrino - tungsten interactions at the LHC energies and estimate the associated pion momentum,  energy and angular distributions. The contribution of the $F_5$ is also investigated. Our results indicate that the pion properties are sensitive to the tau polarization state as well as to the magnitude of $F_5$.
\end{abstract}



\maketitle

\vspace{1cm}

\section{Introduction}

The tau neutrino is commonly known as the least studied fermion in the Standard Model. Its first experimental evidences date back to the beginning of the XXI century, when  different experiments have reported measurements of this neutrino flavor, such as DONuT \cite{DONUT:2000fbd,DONuT:2007bsg} and OPERA \cite{OPERA:2018nar}, which use accelerators as sources of muonic neutrinos that convert into tauonic neutrinos through the oscillation mechanism. In addition,  atmospheric tau neutrinos and from astrophysical sources have been measured by SuperKamiokande \cite{Super-Kamiokande:2017edb} and IceCube Observatory \cite{IceCube:2019dqi,IceCube:2020fpi,IceCube:2024nhk}. These measurements are on the TeV-PeV neutrino energy scale in the case of the measurements at the IceCube, and at few GeV in the other experiments.  Recently, the first muon neutrino from colliders were detected in the FASER \cite{FASER:2021mtu,FASER:2023zcr} and SND@LHC \cite{SNDLHC:2023pun} experiments at the Large Hadron Collider (LHC). Currently, there is the expectation that the study of  tau neutrino physics in the GeV-TeV range will be feasible in the future Forward Physics Facility (FPF) \cite{Feng:2022inv}.  During the high luminosity era of LHC, between 2000 and 20000 tauonic neutrino events are expected at FASER$\nu$2 at the FPF, depending on the hadronic physics model used to simulate the production of frontal hadrons at the LHC \cite{Feng:2022inv}. This expected high number of events will allow testing properties of the Standard Model as well as searching for physics beyond the Standard Model  \cite{Candido:2023utz,Cruz-Martinez:2023sdv}.

\begin{figure}[t]
\includegraphics[scale=0.25]{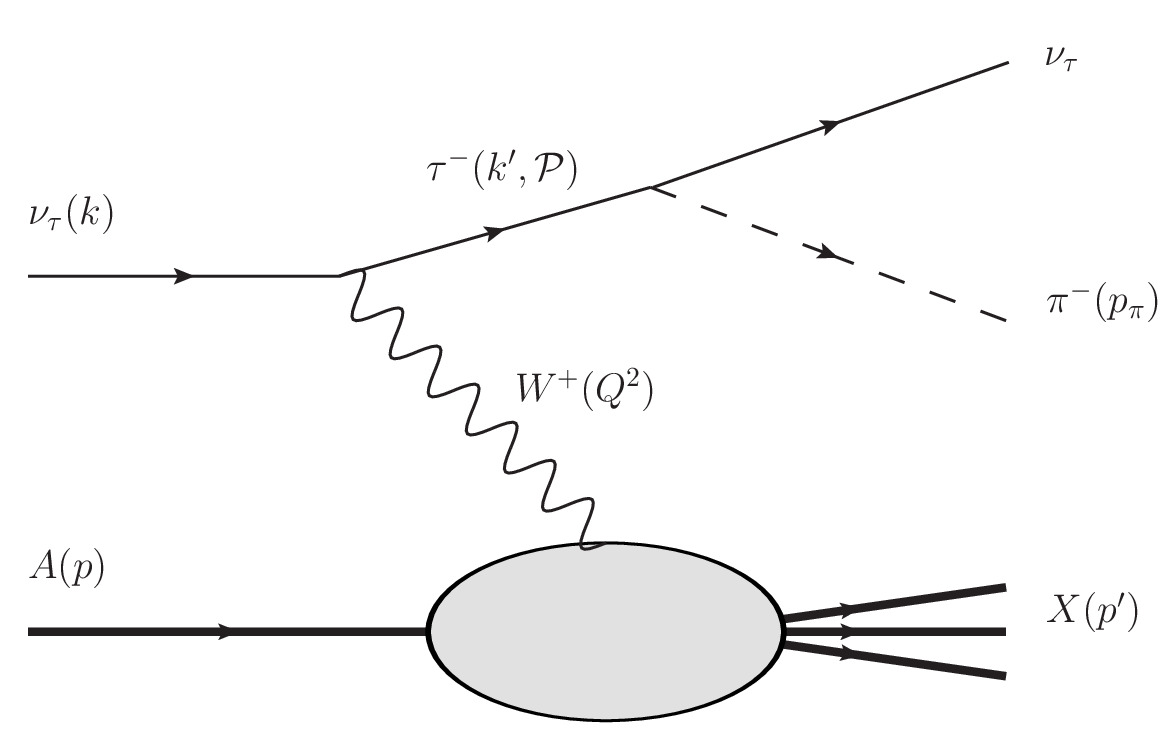} 
\caption{Production of a tau lepton with momentum $k^{\prime}$ and polarization ${\cal{P}}$ in a charged current (CC) deep inelastic  $\nu_{\tau} A$ scattering and its further decay into a pion with momentum $p_\pi$.}
\label{Fig:diagram}
\end{figure}

In the last decades, several works have highlighted the possibility to measure tau polarization observables produced in charged current tau neutrino interactions  considering the current and forthcoming experiments (See, e.g., Refs. \cite{Hagiwara:2003di,Hernandez:2022nmp,Zaidi:2023hdd,Isaacson:2023gwp}). The pioneering works on this topic have investigated the tau polarization considering neutrino interactions in  an energy range slightly above the mass of the tau produced or in the ultra-high energy regime ($E_{\nu} \gtrsim 100$ TeV), where the tau is fully polarized. On the other hand, the analysis of the GeV - TeV neutrino energy range, which will covered by FASER$\nu$2 at FPF, was performed in Ref. \cite{Francener:2024ney}, which have demonstrated that the tau lepton is not fully polarized for these neutrino energies. Such a result motivate us to investigate the impact of the tau polarization on the charged particles that arise from its decay, which will be the particles that will be measured, since the tau decay occurs almost promptly. Our goal in this paper is to complement the study performed in Ref.  \cite{Francener:2024ney},  which have obtained the degree of tau polarization and its longitudinal and transverse components in the FASER$\nu$2 energy regime, investigating how the polarization impacts the momentum, energy and angular distributions of the charged particles arising from the tau decay.  In particular, we will investigate the process represented in Fig. \ref{Fig:diagram}, in which the tau produced in the charged current tau neutrino - nucleus interaction decays into a neutrino plus a charged pion. In our study, we will consider the nucleus target is a tungsten, as expected in the FASER$\nu$2, and estimate the distributions considering different values for the energy of the incoming neutrino. In addition, we also will investigate the impact of the $F_5$ structure function on our predictions. Such an analysis is motivated by the fact that this quantity has not yet been measured in this energy range and, therefore, it is important to search for observables that are sensitive to its magnitude.

This paper is organized as follows. In the next section we present a brief review of the formalism used to estimate the momentum, energy and angular distributions of the charged pion produced in the tau decay. In addition, the expressions for the  longitudinal and transverse components of the tau polarization are presented. 
In Section \ref{sec:res} we present our results for the pion differential distributions considering tau  neutrino - tungsten interactions at different energies of the incoming neutrino.  Finally, in Section \ref{sec:sum}, we summarize our results and conclusions.

\section{Formalism}
\label{sec:for}

In our analysis, we will investigate the impact of polarization $\cal{P}$ of a tau lepton produced in a charged current deep inelastic $\nu_{\tau} A$ scattering  on the longitudinal and transverse momentum of pion that comes from tau decay. The  process is represented in Fig. \ref{Fig:diagram}.  In the laboratory frame, one has that a tau neutrino with four-momentum $k^\mu = (E_\nu , 0, 0, E_\nu)$ collides with the nucleus target of four-momentum $p^\mu$ and mass $M_A$, becoming a tau of four-momentum $k'^{\mu}$ scattered at an angle $\theta_\tau$ relative to the neutrino incidence axis. In the interaction, a $W^{+}$ - boson with square four-momentum $q^2 \equiv - Q^2 = (k - k^{\prime})^2$ and mass $M_W$ is exchanged. The target nucleus becomes a final state of many particles, $X$, characterized by invariant mass $W^2 = (p + q)^2$. The deep inelastic scattering is characterized by $Q^{2}>1\,$GeV and $W>1.4\,$GeV.
As  the tau lifetime is of the order of $10^{-13}$ s, its decay occurs almost promptly, which makes the measurement of its momentum and polarization a very hard task. However, the  tau polarization  affects the observables associated with the particles produced in its decay, as demonstrated, e.g., in Ref. \cite{Hernandez:2022nmp}. In what follows, we will present a brief review of the formalism proposed in Ref. \cite{Hernandez:2022nmp}. We will present the equations for the decay of tau into a tau neutrino plus a charged pion, but the same approach can also be used to estimate the distributions associated with the tau decay into a tau neutrino plus a $\rho$ meson, simply by changing the meson mass. Our focus will be in the production of a pion,  since the effects of tau polarization become more pronounced in this final state \cite{Hernandez:2022nmp}. One has that the double differential cross-section, with respect to the pion energy ($E_\pi$) and the cosine of the scattering angle ($\theta_\pi$) of the pion with respect to the initial neutrino momentum, is given by
\begin{eqnarray}
    \begin{aligned}
    \frac{\mathrm{d}^2\sigma_A^{\pi}}{\mathrm{d}E_\pi\mathrm{d\,cos}\,\theta_\pi} = 
    {\cal{B}_{\pi}} 
    \frac{m_\tau^{2}}{m_\tau^{2}-m_\pi^{2}}
    \frac{G_F^{2} M_N}{2 \pi^{2}}
    \int_{E_{\tau}^{-}}^{E_{\tau}^{\mathrm{sup}}}\mathrm{d}E_{\tau}
    \int_{\mathrm{cos}\,(\theta_{\pi}+\theta_{\tau\pi})}^{\mathrm{cos}\,(\theta_{\pi}-\theta_{\tau\pi})}
    \frac{\mathrm{d}(\mathrm{cos}\,\theta_\tau)F(E_\tau, \mathrm{cos}\,\theta_\tau)}
    {\sqrt{[\mathrm{cos}\,(\theta_\pi-\theta_{\tau\pi})-\mathrm{cos}\,\theta_\tau]
    [\mathrm{cos}\,\theta_\tau-\mathrm{cos}\,(\theta_\pi-\theta_{\tau\pi})]}} \\
    \left\{ 
    1+\frac{2m_\tau}{m_\tau^{2}-m_\pi^{2}} 
    \frac{m_\tau^{2}-2 m_\pi^{2}}{m_\tau^{2}+2 m_\pi^{2}}
    \left[
    P_L(E_\tau, \mathrm{cos}\,\theta_\tau)
    \left(
    \frac{E_\pi|\vec{k}'|}{m_\tau} - \frac{E_\tau|\vec{p}_\pi|}{m_\tau}\mathrm{cos}\,\theta_\pi 
    \right) 
    \right. \right.
    \\
    \left.
    \left.
    + P_T(E_\tau, \mathrm{cos}\,\theta_\tau)
    \frac{|\vec{p}_\pi|(\mathrm{cos}_\pi-\mathrm{cos}\,\theta_\tau\mathrm{cos}\,\theta_{\tau\pi}))}{\mathrm{sin}\,\theta_\tau}
    \right] 
    \right\}\, ,
    \label{eq:sigma1}
    \end{aligned}
\end{eqnarray}
where ${\cal{B}}_\pi$ is the branching fraction of tau lepton decay in tau neutrino plus pion, $\vec{p}_\pi$ and $m_\pi$ the tri-momentum and mass of the pion, respectively, and $\theta_{\tau\pi}$ the angle between $\vec{p}_\pi$ and $\vec{k}'$, defined by
\begin{eqnarray}
    \mathrm{cos}\, \theta_{\tau\pi} = 
    \frac{2E_\tau E_\pi -m_\tau^{2}-m_\pi^{2}}{2|\vec{k}'||\vec{p}_\pi|}
    \, .
    \label{eq:cos_theta_tau_pion}
\end{eqnarray}
{ For the incident neutrino energy greater than $(m_{\tau}^{2}+m_{\pi}^{2})/2m_{\pi}$, the limits allowed for the pion and tau energy are given by}
\begin{eqnarray}
    \begin{aligned}
    m_{\pi} \leq E_{\pi} \leq E_{\pi}^{\mathrm{max}}
     , \,\,\,\,\,
    E_{\tau}^{\mathrm{sup}} = 
    \Theta(E_{\pi}^{\mathrm{int}} - E_{\pi})E_{\tau}^{+} 
    + \Theta(E_{\pi} - E_{\pi}^{\mathrm{int}})E_{\nu}
    \label{eq:limites2}
    \end{aligned}
\end{eqnarray}
{ where $\Theta$ is the step function, and  we introduce the auxiliary variables }
\begin{eqnarray}
    \begin{aligned}
    E_{\tau}^{\pm} = 
    \frac{(m_{\tau}^{2}+m_{\pi}^{2})E_{\pi}\pm (m_{\tau}^{2}-m_{\pi}^{2})|\vec{p}_{\pi}|}{2m_{\pi}^{2}}
     , \,\,\,\,\,
    E_{\pi}^{\mathrm{max, int}} = 
    \frac{(m_{\tau}^{2}+m_{\pi}^{2})E_{\nu}\pm (m_{\tau}^{2}-m_{\pi}^{2})\sqrt{E_{\nu}^{2}-m_{\tau}^{2}}}{2m_{\tau}^{2}} \, .
    \label{eq:limites1}
    \end{aligned}
\end{eqnarray} 
From the experimental point of view, it is interesting to express the cross-section in terms of the longitudinal ($p_{L\pi}$) and transverse ($p_{T\pi}$) moments of the pion resulting from the tau lepton decay. Such a double differential cross-section is related with Eq. (\ref{eq:sigma1}) by the following expression \cite{Hernandez:2022nmp}
\begin{eqnarray}
    \begin{aligned}
    \frac{\mathrm{d}^2\sigma_A^{\pi}}{\mathrm{d}p_{L\pi}\mathrm{d}p_{T\pi}} = 
    \frac{p_{T\pi}}{(p_{L\pi}^{2}+p_{T\pi}^{2})^{1/2}}
    \frac{1}{(p_{L\pi}^{2}+p_{T\pi}^{2} + m_{\pi}^{2})^{1/2}}
    \left[
    \frac{\mathrm{d}^2\sigma_A^{\pi}}{\mathrm{d}E_\pi\mathrm{d\,cos}\,\theta_\pi}
    \right]
    \, .
    \label{eq:sigma2}
    \end{aligned}
\end{eqnarray}

One has that the differential cross - sections, defined by Eqs. (\ref{eq:sigma1}) and (\ref{eq:sigma2}), are determined by the function $F(E_\tau, \mathrm{cos}\,\theta_\tau)$ and by the longitudinal ($P_L$) and transverse ($P_T$) components of the tau polarization. Such quantities are determined by the structure functions $F_i$ ($i = 1 - 5$) and are given in the laboratory frame by (For details see, e.g., Ref.  \cite{Francener:2024ney})
\begin{eqnarray}
    F(E_\tau, \mathrm{cos}\,\theta_\tau)  =   
    \left\{ 2F^A_{1}(x,Q^2)(E_\tau -|\vec{k}'|\mathrm{cos}\,\theta_\tau) + 
    F^A_{2}(x,Q^2)\frac{M_A}{\nu}(E_\tau+|\vec{k}'|\mathrm{cos}\,\theta_\tau) \right. \nonumber \\
    \pm F^A_{3}(x,Q^2)\frac{1}{\nu}[| \vec{k}' |^2 + E_\nu E_\tau - (E_\nu + E_\tau )|\vec{k}'|\mathrm{cos}\,\theta_\tau] + 
    F^A_{4}(x,Q^2)\frac{m_\tau^{2}}{\nu M_A x} (E_\tau - |\vec{k}'|\mathrm{cos}\,\theta_\tau) + \nonumber \\
    \left. - F^A_{5}(x,Q^2)\frac{2m_\tau^2}{\nu} \right\} \, , 
     \label{eq:F}
\end{eqnarray}
\begin{eqnarray}
    P_L  =  
    \mp \frac{1}{F(E_\tau, \mathrm{cos}\,\theta_\tau)} \left\{ \left[ 2F^A_{1}(x,Q^2) - F^A_{4}(x,Q^2)\frac{m_\tau^{2}}{\nu M_A x} \right] (|\vec{k}'| - E_\tau\mathrm{cos}\,\theta_\tau) + 
    F^A_{2}(x,Q^2)\frac{M_A}{\nu}(|\vec{k}'| + E_\tau\mathrm{cos}\,\theta_\tau) \right. \nonumber \\
    \pm \frac{F^A_{3}(x,Q^2)}{\nu}\frac{1}{\nu}[ (E_\nu + E_\tau )|\vec{k}'| - (| \vec{k}' |^2 + E_\nu E_\tau)\mathrm{cos}\,\theta_\tau] 
    \left. - F^A_{5}(x,Q^2)\frac{2m_\tau^2}{\nu} \mathrm{cos}\,\theta_\tau \right\}
     \label{eq:PL}
\end{eqnarray}
and
\begin{eqnarray}
\begin{aligned}
    P_T = 
    \mp \frac{m_\tau \mathrm{sin}\,\theta_\tau}{F(E_\tau, \mathrm{cos}\,\theta_\tau)} \left[ 2F^A_{1}(x,Q^2) -  F^A_{2}(x,Q^2)\frac{M_A}{\nu} 
    \pm F^A_{3}(x,Q^2) \frac{E_\nu}{\nu} - F^A_{4}(x,Q^2)\frac{m_\tau^{2}}{\nu M_A x} 
     + F^A_{5}(x,Q^2)\frac{2E_\tau}{\nu} \right] \, ,
     \label{eq:PT}
\end{aligned}
\end{eqnarray}
with the upper (lower) sign for (anti) neutrinos, $m_{\tau}$ the mass of the tau lepton and $\nu = E_{\nu} - E_{\tau}$.
It is important to emphasize that $P_T$ is proportional to $\mathrm{sin}\,\theta_\tau$ and $m_\tau$, vanishing for $\theta_\tau = 0^\circ$, being important when the energy of the tau produced is not much greater than its mass. The degree of polarization of the tau lepton produced is defined as $ {\cal{P}} = \sqrt{P_L^2 + P_T^2} $
 which is less than or equal to 1, being 0 when the particle is unpolarized and 1 when it is fully polarized. 
In our analysis, we will assume the validity of the Callan-Gross \cite{Callan:1969uq} and Albright-Jarlskog \cite{Albright:1974ts} relations in the limit studied here, which allow us to write $F^A_{1}$ and $F^A_{5}$ in terms of $F^A_{2}$, respectively. It is important to emphasize that the Albright-Jarlskog relation implies $F_4 = 0$. 
Moreover, the nuclear structure functions will be estimated taking into account of the nuclear effects in the parton distributions (nPDF), as predicted in the nCTEQ15 parametrization \cite{Kovarik:2015cma,Duwentaster:2022kpv,Muzakka:2022wey}.

\begin{figure}[t]
	\centering
	\begin{tabular}{c}
	\includegraphics[width=0.7\textwidth]{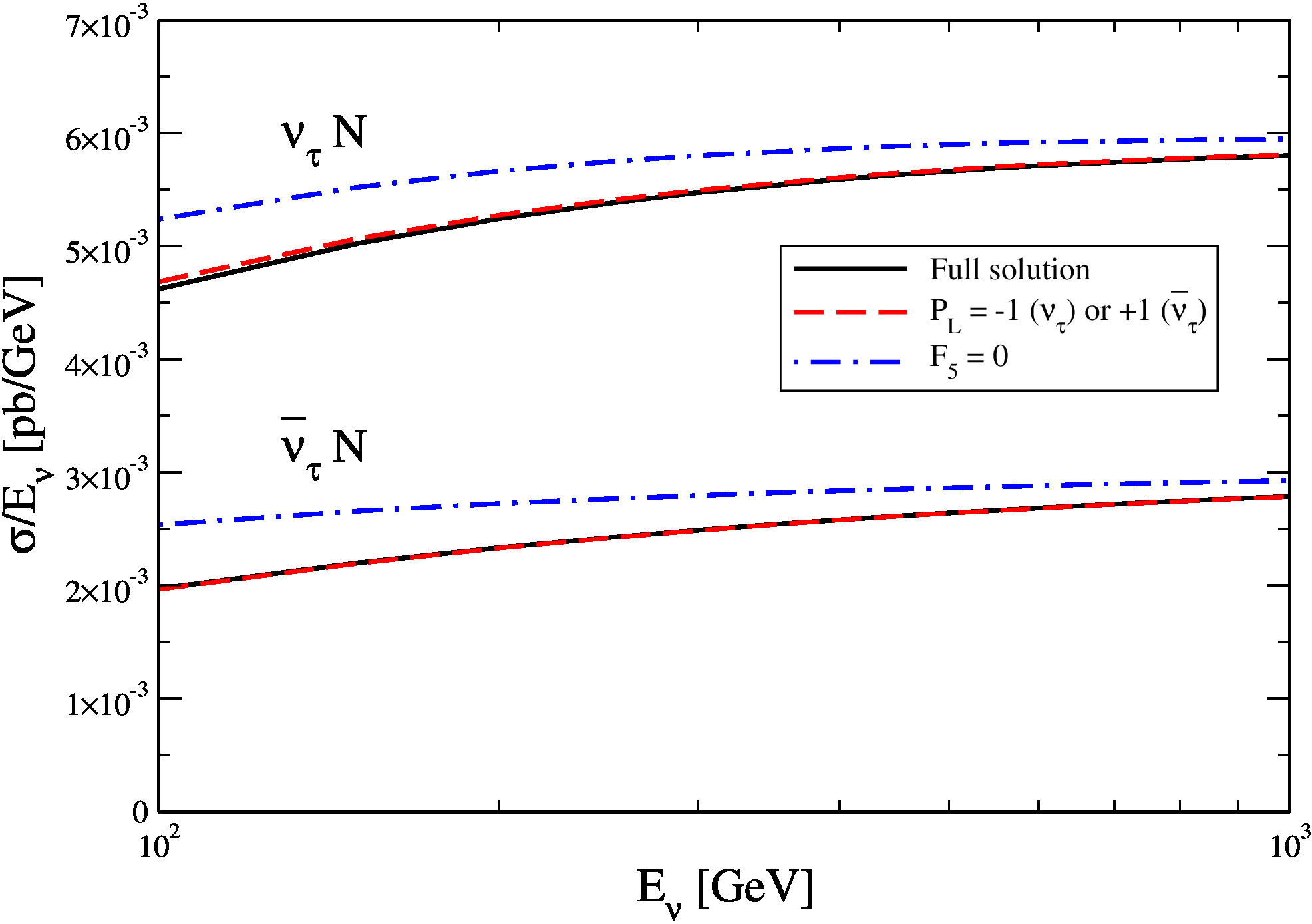} 
	\end{tabular}
\caption{Predictions for the energy dependence of the $\nu_\tau / \bar{\nu}_\tau $ - tungsten cross-sections per nucleon, derived considering the  full solution of Eq. (\ref{eq:sigma1}) [solid lines]. For comparison, the approximated results obtained assuming that the (anti)tau produced is fully polarized (dashed lines) or  assuming that $F_5 = 0$ (dashed - dotted lines) are also presented.}
\label{fig:sigma}
\end{figure}

\section{Results}
\label{sec:res}
Initially,  in Fig. \ref{fig:sigma},  we present the energy dependence of the $\nu_\tau / \bar{\nu}_\tau $ - tungsten cross-sections  per nucleon in the range of 100 GeV to 1000 GeV, which is expected to be probed at the future FPF. The predictions derived using Eq. (\ref{eq:sigma1}), taking into account that the tau lepton is not fully polarized in the considered energy range, are represented by the solid lines. For comparison, we also present the results obtained assuming that the (anti)tau produced is fully polarized (dashed lines).
The results indicate that considering the complete polarization of (anti)tau does not change significantly the cross-section, especially in processes induced by antineutrinos. In addition, we show the predictions derived assuming that $F_5 = 0$ (dashed - dotted lines). Such an analysis is motivated by the fact that the tau mass in the order of GeV, which implies a non-negligible contribution of the structure function $F_5$, which multiplies $m_\tau^{2}$ in Eq. (\ref{eq:F}). Moreover, one has that this structure function has never been measured in this energy range.
The results presented in Fig. \ref{fig:sigma} indicate that if the contribution  of the $F_5$ structure function is neglected, the predictions are enhanced. In particular, if $F_5$ is assumed to be equal to zero, the cross-sections increase by 13.4\% (28.4\%) and 2.4\% (5.0\%) for reactions induced by (anti)neutrinos of energy equal to 100 GeV and 1000 GeV, respectively. These results are similar to those obtained in Ref. \cite{Jeong:2010nt}.

\begin{figure}[t]
	\centering
	\begin{tabular}{ccc}
    \includegraphics[width=0.5\textwidth]{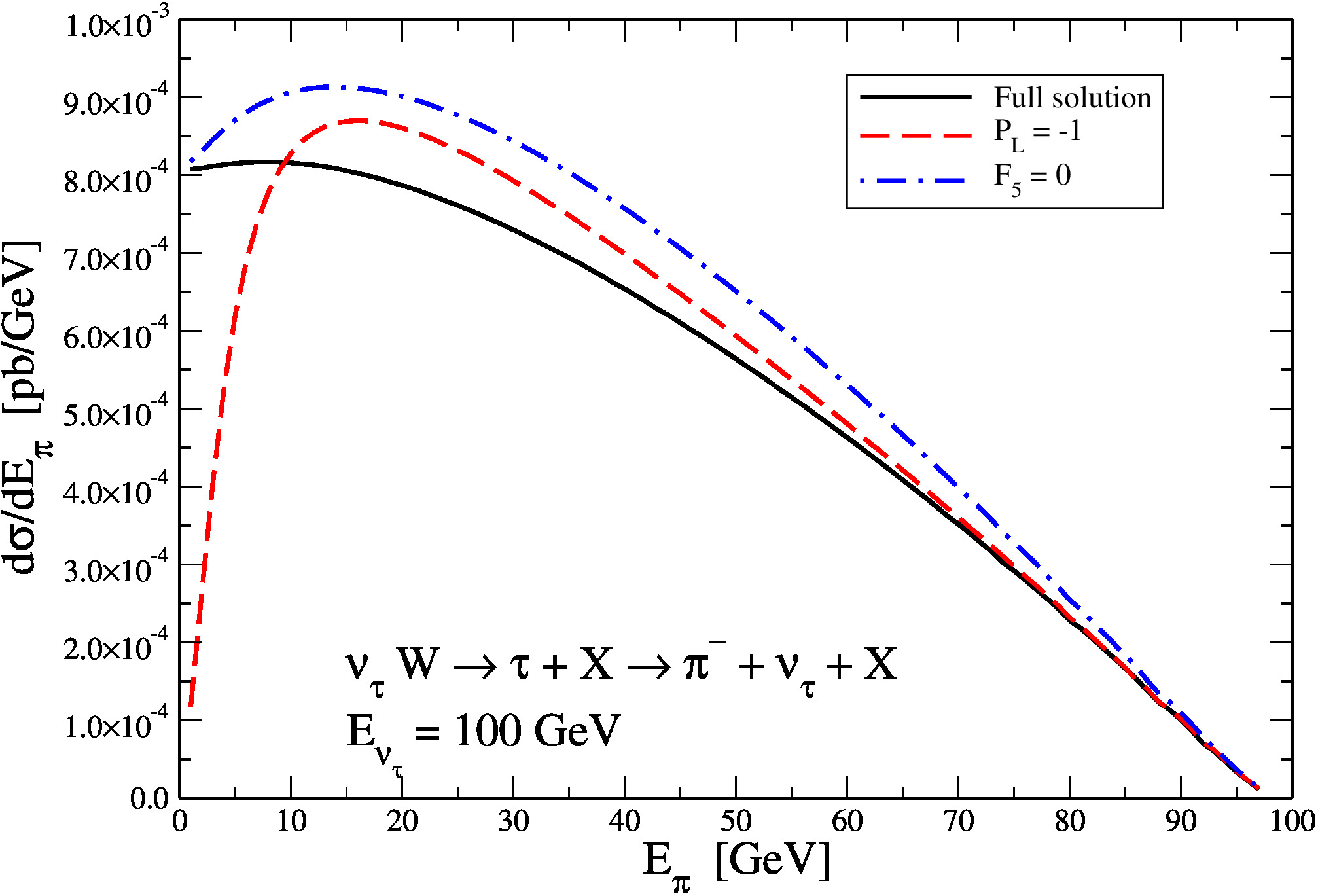} &
    \includegraphics[width=0.5\textwidth]{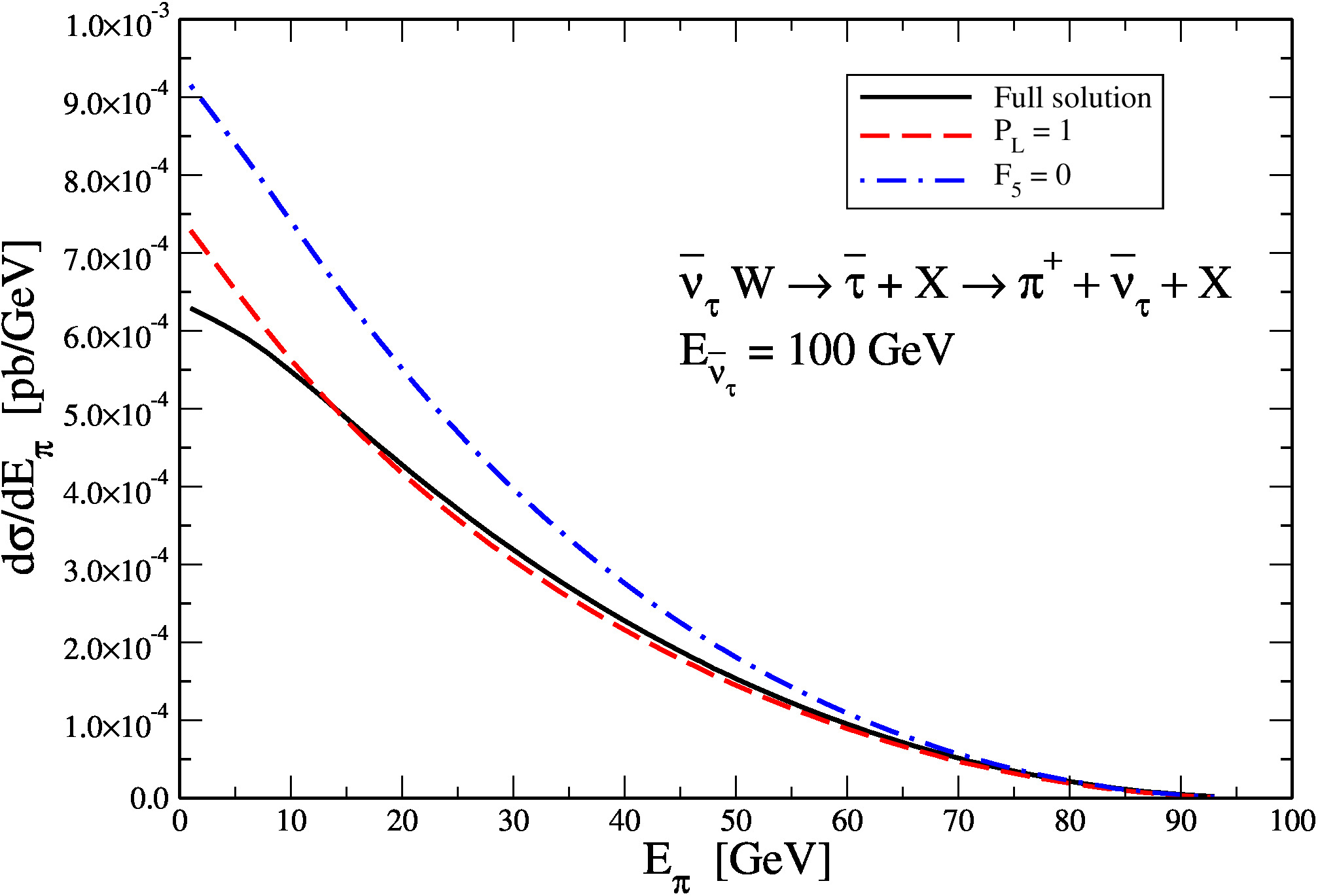} \\
    \includegraphics[width=0.5\textwidth]{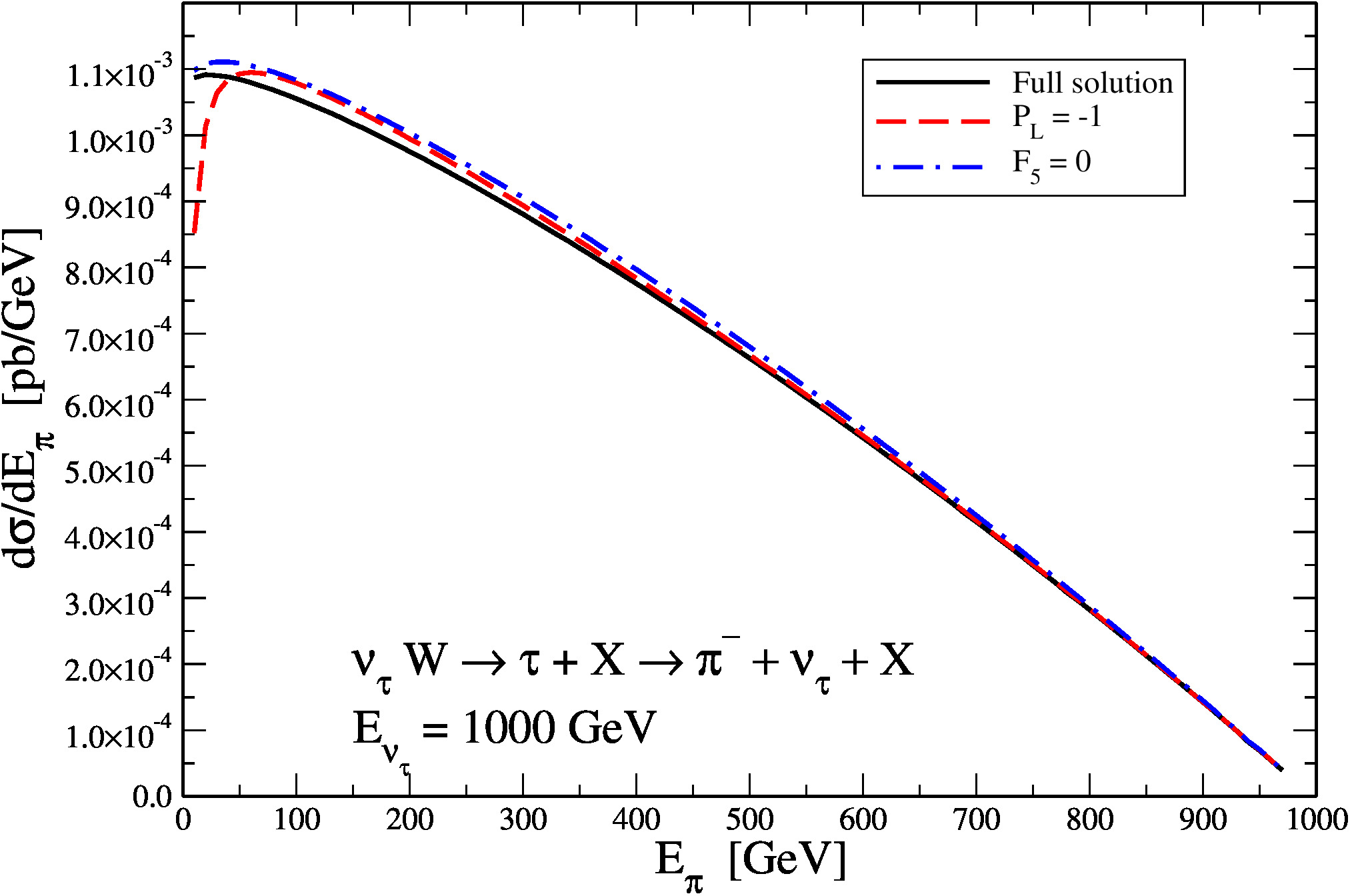} &
    \includegraphics[width=0.5\textwidth]{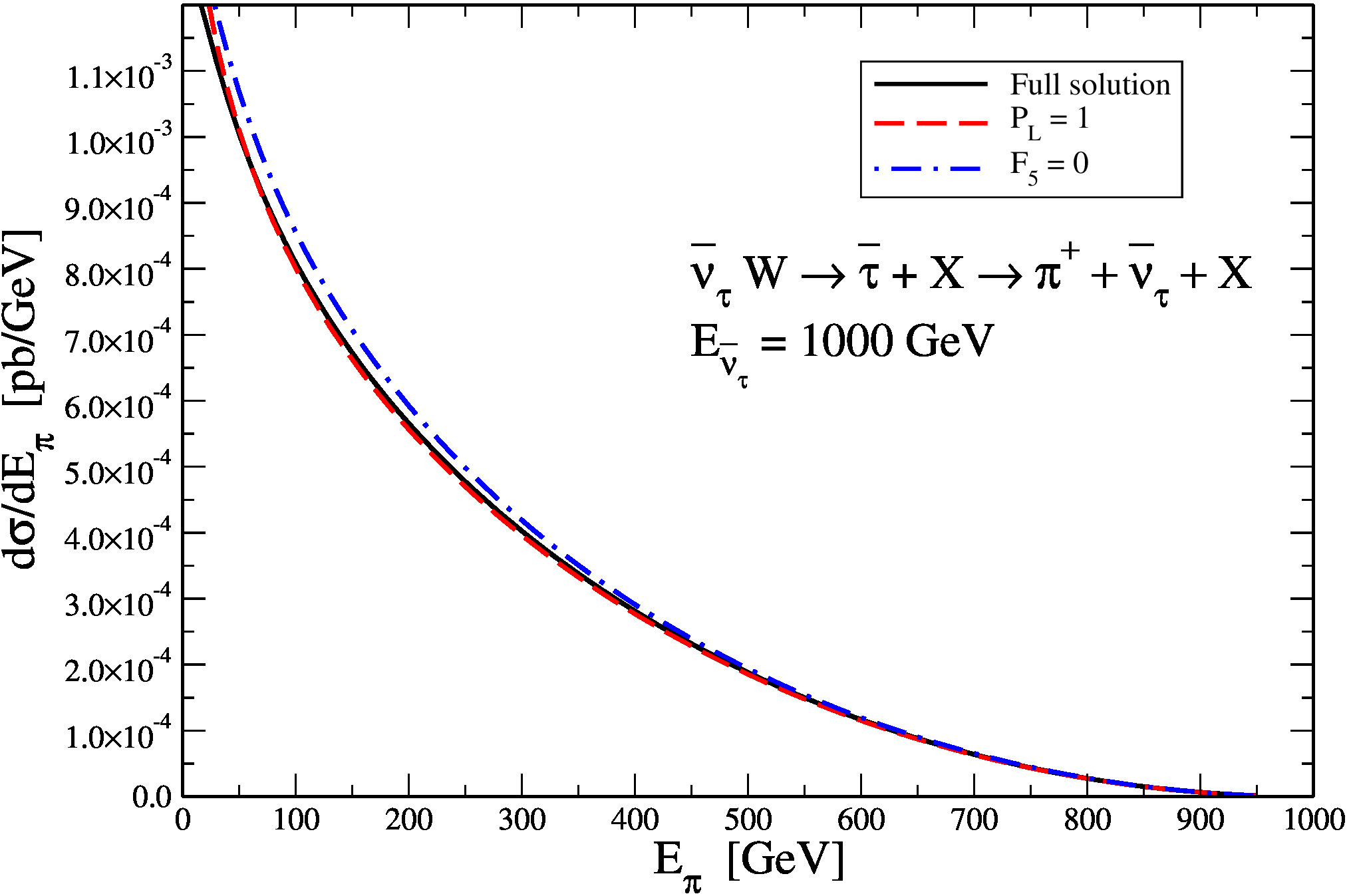} 
			\end{tabular}
\caption{ Differential cross-section with respect to pion energy ($E_\pi$) for the  $ \nu_\tau + W \rightarrow \tau^{-} + X \rightarrow \pi^{-} + \nu_\tau  + X$ (left panels)
and $ \bar{\nu}_\tau + W \rightarrow \tau^{+} + X \rightarrow \pi^{+}+\bar{\nu}_\tau + X$
(right panels) processes. Predictions for an incoming  (anti)neutrino energy of 100 GeV (upper panels) and 1000 GeV (lower panels). }
\label{fig:dsdE}
\end{figure}

\begin{figure}[t]
	\centering
	\begin{tabular}{ccc}
    \includegraphics[width=0.5\textwidth]{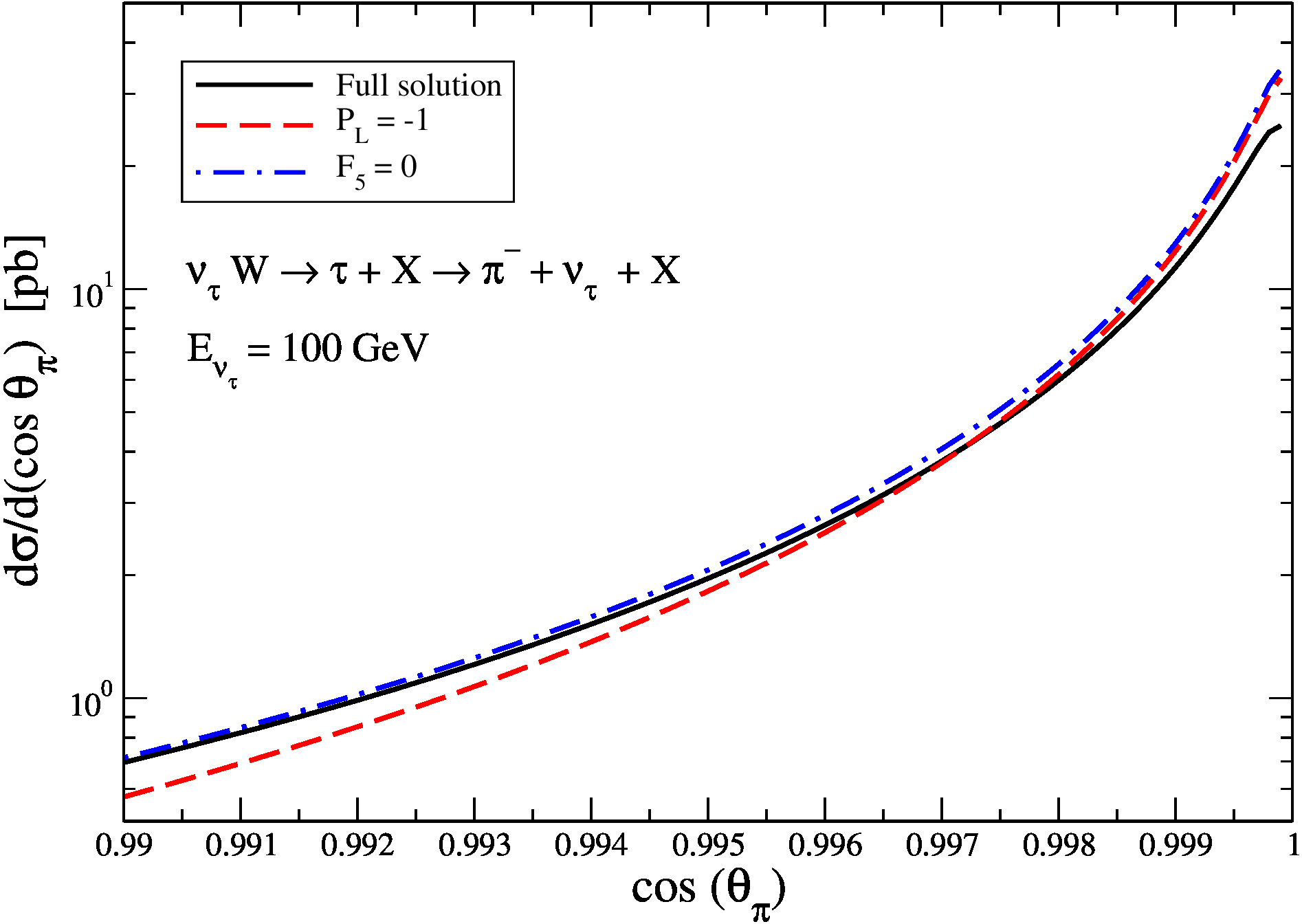} &
    \includegraphics[width=0.5\textwidth]{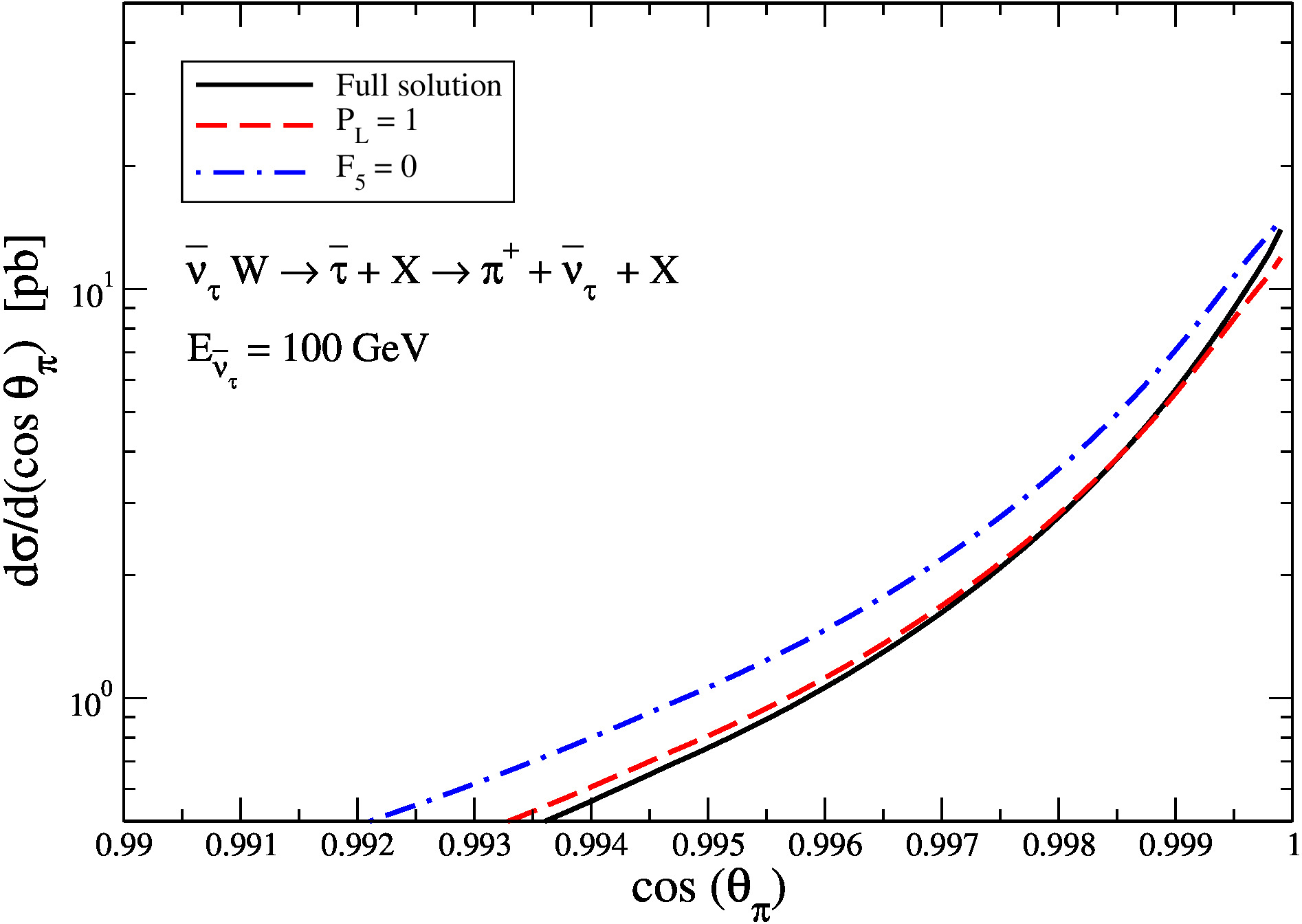} \\
    \includegraphics[width=0.5\textwidth]{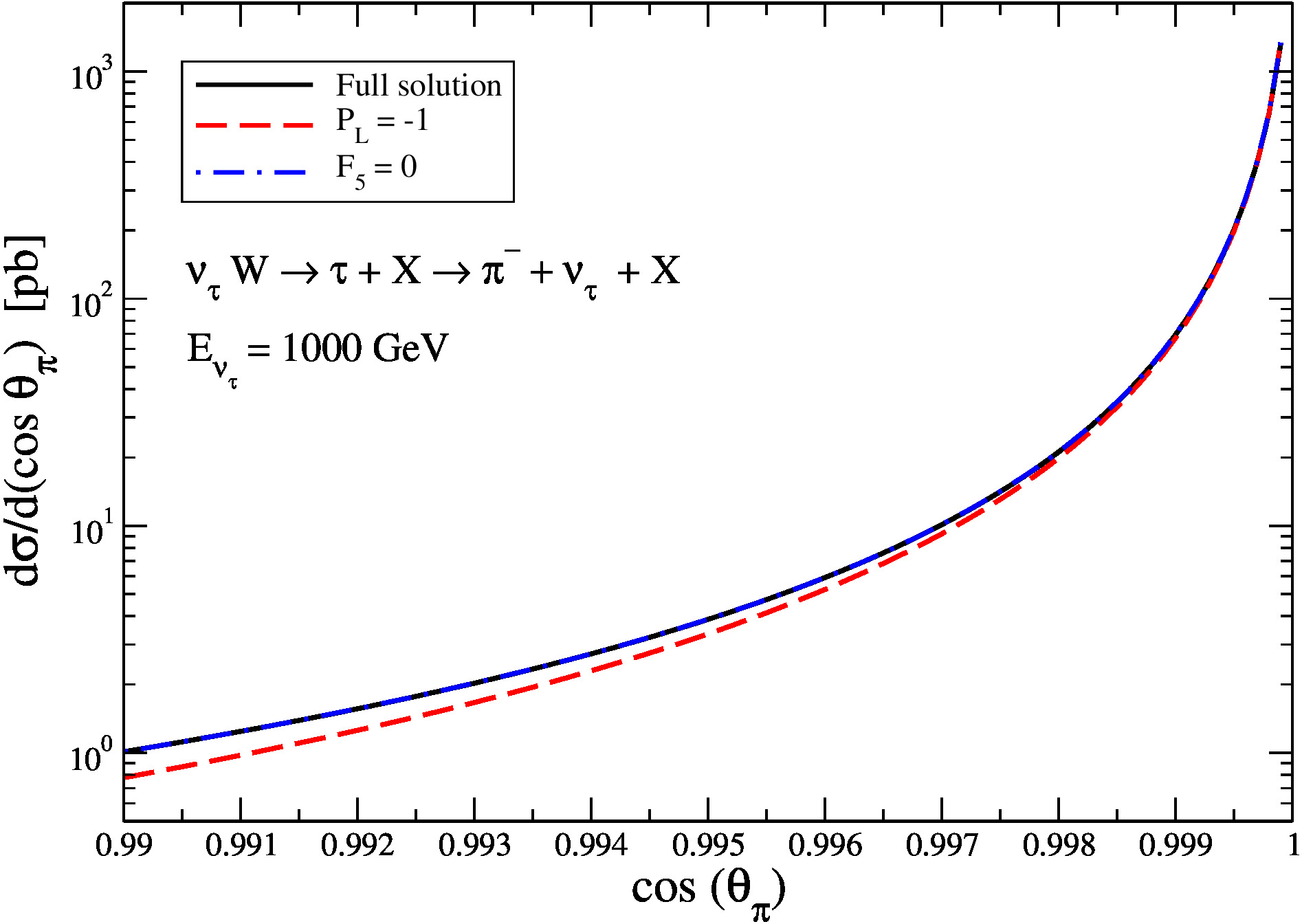} &
    \includegraphics[width=0.5\textwidth]{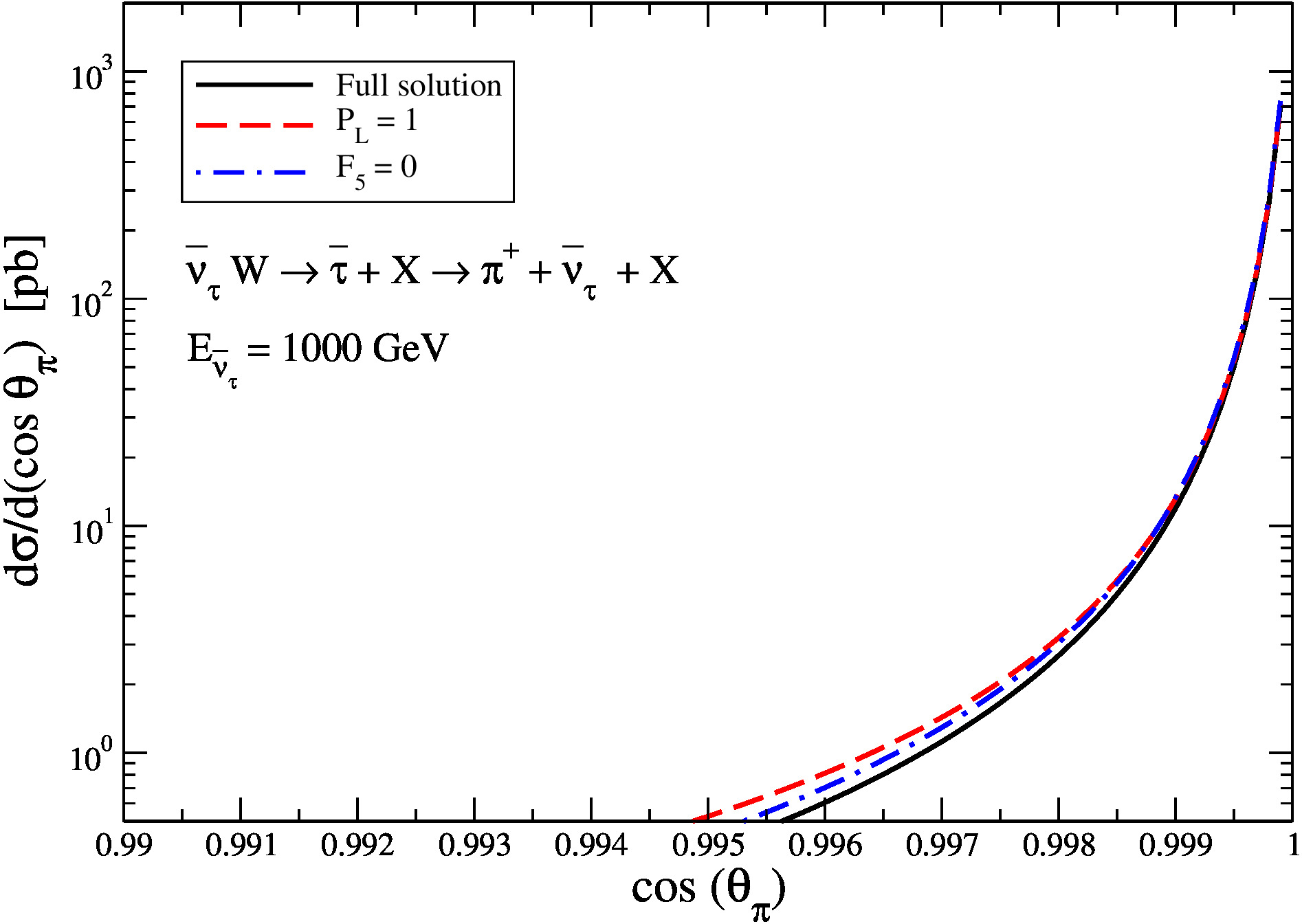} 
			\end{tabular}
\caption{Differential cross-section with respect to cosine of pion scattering angle ($\theta_\pi$) for the $ \nu_\tau + W \rightarrow \tau^{-} + X \rightarrow \pi^{-} + \nu_\tau  + X$ (left panels)
and $ \bar{\nu}_\tau + W \rightarrow \tau^{+} + X \rightarrow \pi^{+}+\bar{\nu}_\tau + X$
(right panels) processes. Predictions for an incoming  (anti)neutrino energy of 100 GeV (upper panels) and 1000 GeV (lower panels). }
\label{fig:dsdcostd}
\end{figure}

In what follows, we will present our results for the pion differential distributions considering 
$\nu_\tau / \bar{\nu}_\tau $ - tungsten interactions at the LHC energy regime. Our goal is to investigate the impact of the tau polarization and the $F_5$ structure function on these distributions. 
 In Figs. \ref{fig:dsdE} and \ref{fig:dsdcostd} we show the differential cross-sections in the pion energy ($E_\pi$) and the cosine of its scattering angle with respect to incident neutrino axis ($\theta_\pi$), respectively. The results are presented for an incoming tau neutrino (left panels) and tau antineutrino (right panels) considering an energy of 100 GeV (upper panels) and 1000 GeV (lower panels). 
 The predictions for $\mathrm{d}\sigma/\mathrm{d}(\mathrm{cos}\, \theta_\pi)$  are presented only for cos $(\theta_\pi) > 0.99$, since  the cross-section is almost entirely contained within this limit, and the future FASER$\nu$2 detector is being designed to measure  particles scattered at small angles, with mrad accuracy.   From Fig. \ref{fig:dsdE} one has that if the tau is assumed completely polarized and the contribution of $F_5$ is disregarded, the predictions are mainly impacted on the region of lower values of the pion energy and, in addition, they decrease with the increasing of the (anti)neutrino energy. The impact of $F_5 = 0$ increases the differential cross-section in almost the entire range of $E_\pi$ for both neutrinos and antineutrinos, demonstrating that these distributions are sensitive to the magnitude of $F_5$. If the tau is assumed to be fully polarized, $\mathrm{d}\sigma/\mathrm{d}E_\pi$ associated with  the (anti)tau decay is smaller (larger) than that obtained with the full solution of Eq. (\ref{eq:sigma1}) at small pion energies. The behavior is reversed when the pion reaches energies of approximately 10 GeV and 50 GeV for incident (anti)neutrinos of 100 GeV and 1000 GeV, respectively. 
 In particular, for an incident (anti) neutrino with energy of 100 GeV, one has that when the (anti) tau produced is  assumed to be fully polarized, the differential cross-section $d\sigma/dE_{\pi}$ for $E_{\pi} = 2$ GeV decreases (increases) by $67\%$ ($14\%$). Such a result demonstrate that the assumption that the tau is fully polarized, usually present in the Monte Carlo generators for $\nu A$ interactions at high energies, is not a good approximation at the LHC energies. 
 
In Fig. \ref{fig:dsdcostd},  we observe a significant impact on the angular distribution of the cross-section for the pion resulting from tau decay if  $F_5$ is assumed to be equal to zero or that  tau produced is completely polarized. Ignoring the contribution of $F_5$, the cross-section induced by (anti)neutrinos is more impacted at smaller (larger) pion scattering angles. On the other hand, the effects of considering the tau as completely polarized decrease (increase) the cross-section at larger angles and increase (decrease) it at smaller scattering angles for pions resulting from (anti)tau. We also observe that the differential cross-section  exhibits more significant changes at lower incident neutrino energies.
  Assuming that the (anti) tau is fully polarized and  an incident (anti) neutrino with energy of 100 GeV, one has verified that the differential cross-section in the pion scattering angle decreases (increases) by $7.0\%$ ($6.9\%$) and increases (decreases) by $30\%$ ($15\%$) considering $\mathrm{cos}\,\theta_\pi = 0.995$ and $\mathrm{cos}\,\theta_\pi = 0.9999$, respectively. For an incident (anti) neutrino of 1000 GeV, such observable decreases (increases) $14\%$ ($40\%$) for $\mathrm{cos}\,\theta_\pi = 0.995$ and increases (decreases) by $1.2\%$ ($0.2\%$) for $\mathrm{cos}\,\theta_\pi = 0.9999$. Such a result confirms that in order to derive a realistic prediction of the pion production in $\nu W$ interactions at the LHC energies, we have to take into account that the tau is not fully polarized.

\begin{figure}[t]
	\centering
	\begin{tabular}{cc}
    \includegraphics[width=0.5\textwidth]{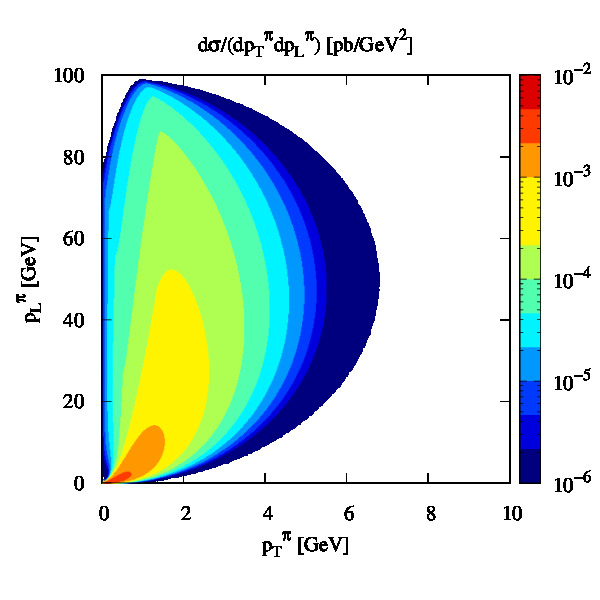} &
    \includegraphics[width=0.5\textwidth]{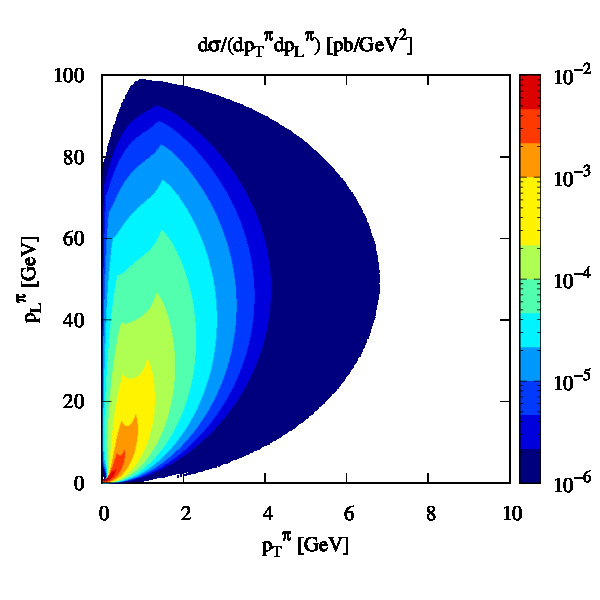} \\
    \includegraphics[width=0.5\textwidth]{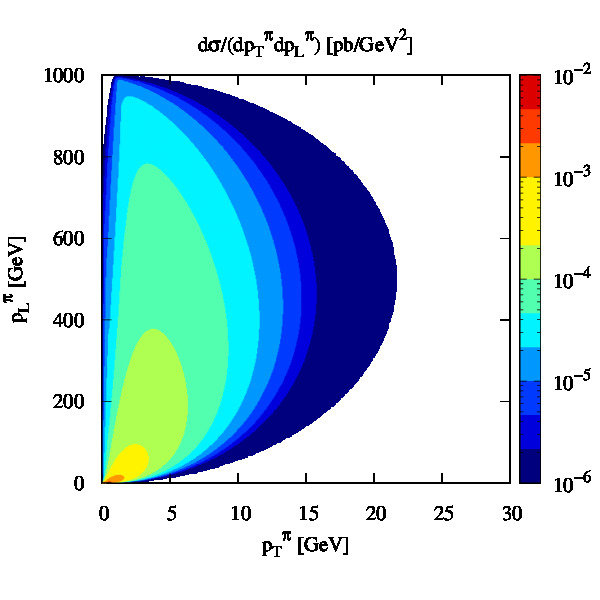} &
    \includegraphics[width=0.5\textwidth]{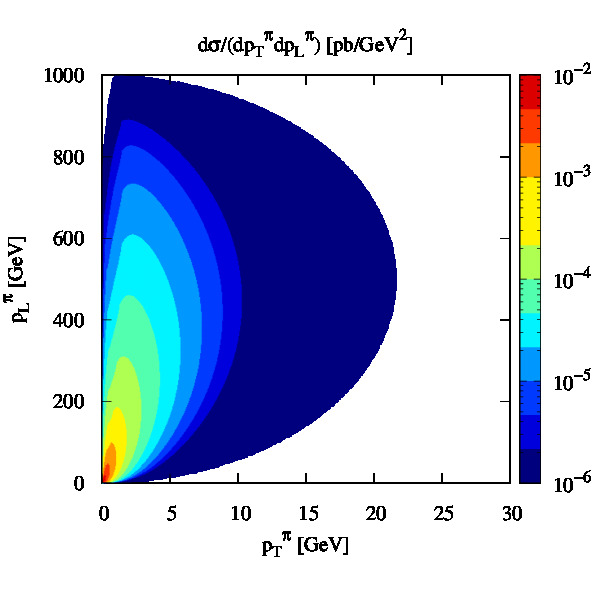}
			\end{tabular}
\caption{ Double differential cross-section for the $\nu_\tau + W \rightarrow \tau^{-} + X \rightarrow \pi^{-}+\nu_\tau + X$ (left panels) and $\bar{\nu}_\tau + W \rightarrow \tau^+ + X \rightarrow \pi^{+}+\bar{\nu}_\tau + X$ (right panels)  processes with respect to longitudinal and transverse pion momentum. Tau neutrino energy is fixed at 100 GeV (upper panels) and 1000 GeV (lower panels). }
\label{fig:dsdptdpl}
\end{figure}

In Fig. \ref{fig:dsdptdpl}, we present our predictions for the double differential cross-sections with respect to the pion's longitudinal and transverse momentum for a tungsten target induced by tau (anti)neutrinos. The results are shown for an incoming tau neutrino (left panels) and tau antineutrino  (right panels) assuming  a neutrino energy of 100 GeV (upper panels)  and 1000 GeV (lower panels). 
{One has  that the differential cross-section induced by neutrinos  is larger  in the region of low longitudinal and transverse momenta of the pion, with the phase space increasing with the neutrino energy. However, in interactions initiated by antineutrinos, the decreasing of the cross-section  at high transverse momentum of the pion is faster in comparison to that predicted for $\nu W$ interactions in all available phase space. }
 In the case of antineutrino induced interactions, especially for an energy of 100 GeV (Fig. \ref{fig:dsdptdpl}, top right), we observe a `dip' in the double differential cross-section at approximately $p_{T}^{\pi} = 0.6$ GeV across the entire range of the pion's longitudinal momentum. This effect arises from the assumed kinematic cuts that characterize DIS: $Q > 1.0$ GeV and $W > 1.4$ GeV. The maximum of the cross-section distribution for neutrino-induced events occurs at larger transverse momentum of the pion compared to antineutrino interactions. Therefore, combined with the greater longitudinal momentum of the pion in neutrino processes, the effect of these cuts is less pronounced for neutrino interactions. 

\begin{figure}[t]
	\centering
	\begin{tabular}{cc}
    \includegraphics[width=0.5\textwidth]{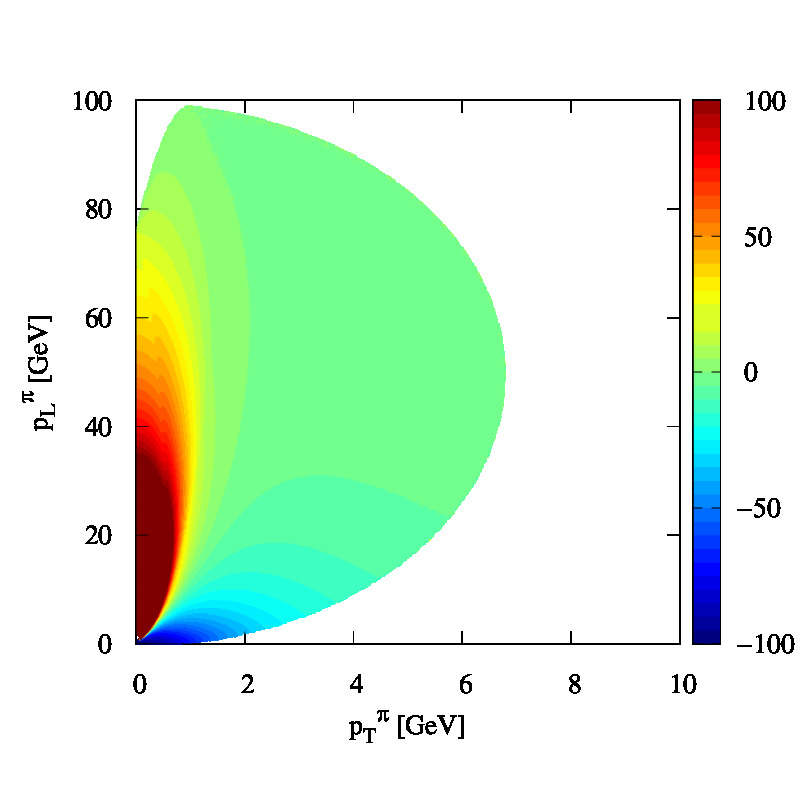} &
    \includegraphics[width=0.5\textwidth]{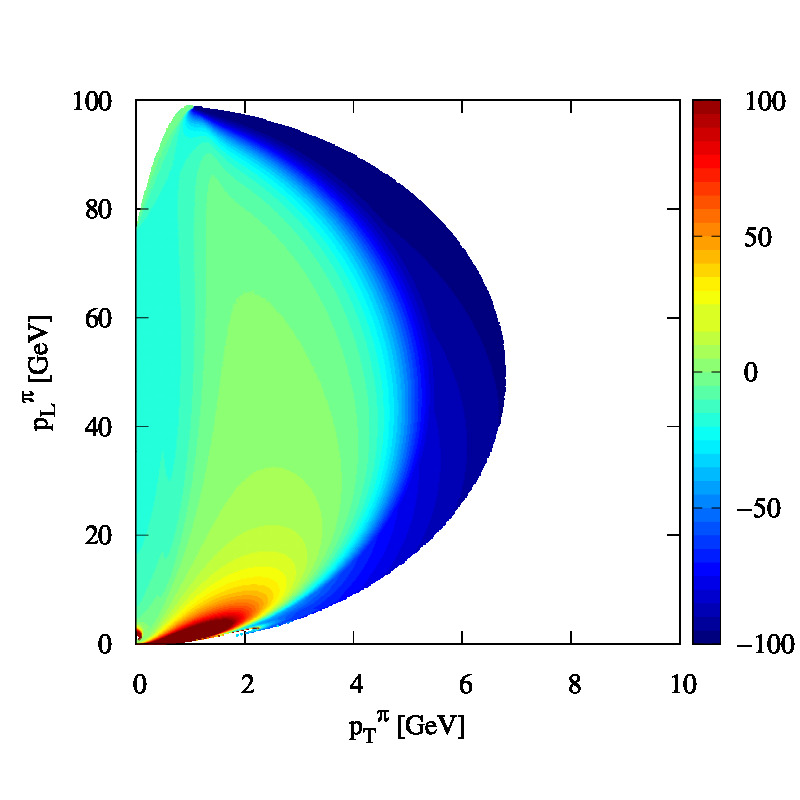} \\
    \includegraphics[width=0.5\textwidth]{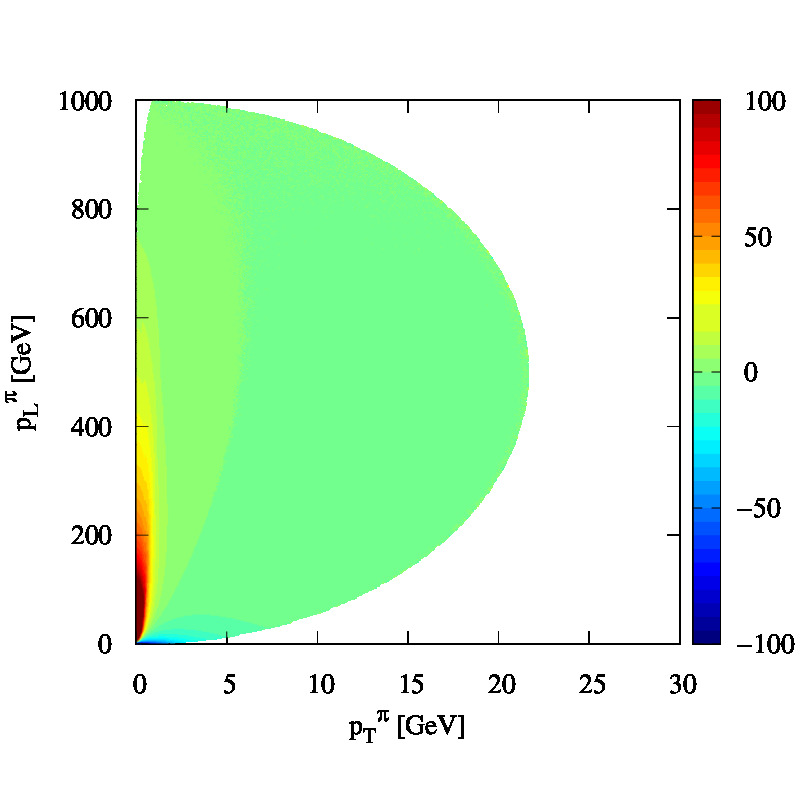} &
    \includegraphics[width=0.5\textwidth]{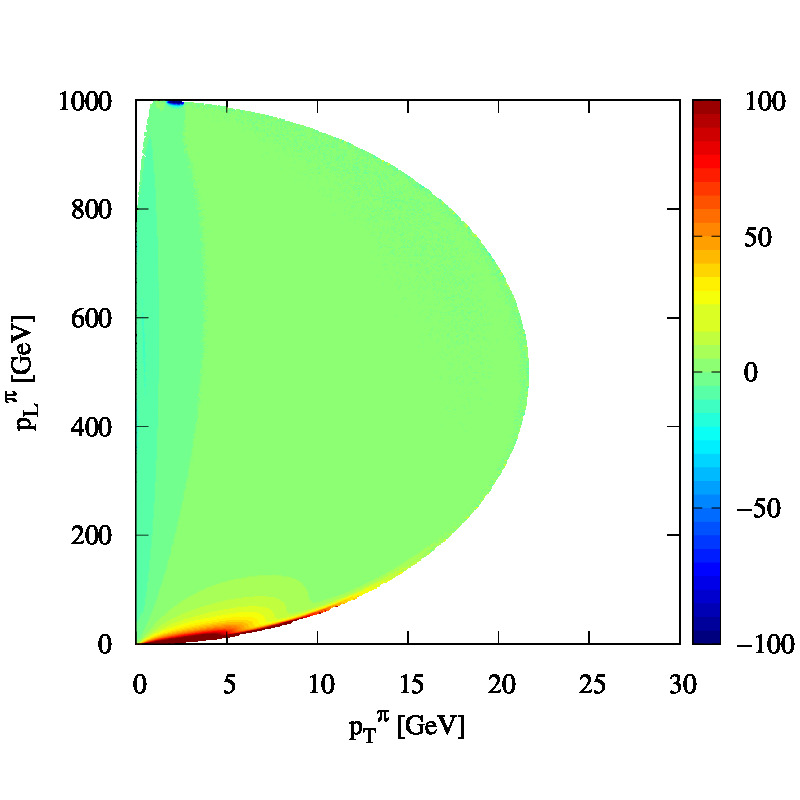}
			\end{tabular}
\caption{ Difference in \% between the predictions for $\mathrm{d}\sigma/(\mathrm{d}p_T^\pi \mathrm{d}p_L^\pi)$ derived assuming the full solution and obtained considering that the (anti)tau produced is full longitudinally polarized in the left (right). Results for an incident (anti)neutrino energy of 100 GeV (upper panels) and 1000 GeV (lower panels).  }
\label{fig:diff_PL1}
\end{figure}

\begin{figure}[t]
	\centering
	\begin{tabular}{cc}
    \includegraphics[width=0.5\textwidth]{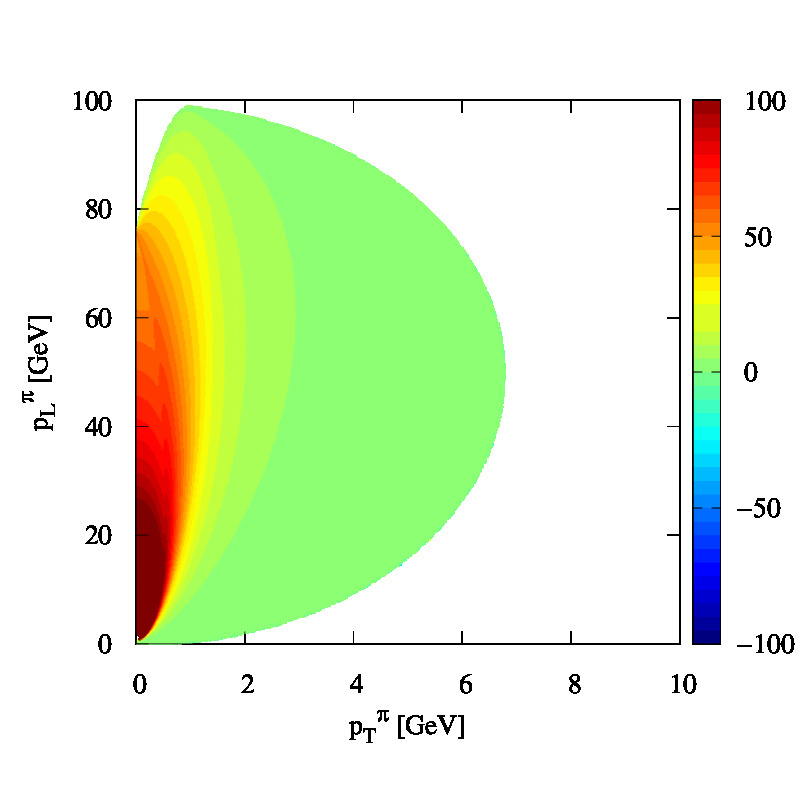} &
    \includegraphics[width=0.5\textwidth]{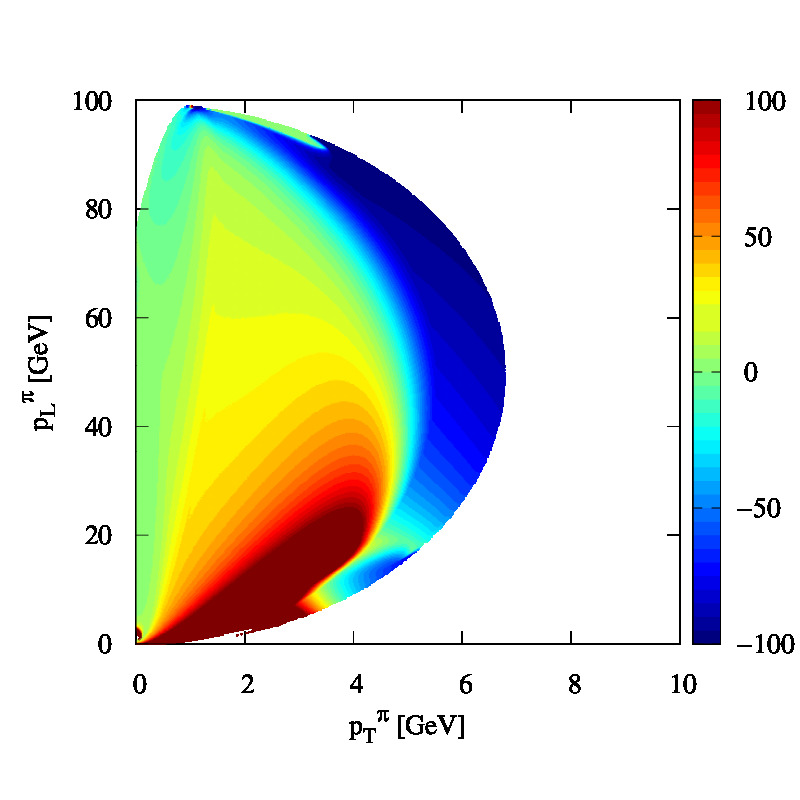} \\
    \includegraphics[width=0.5\textwidth]{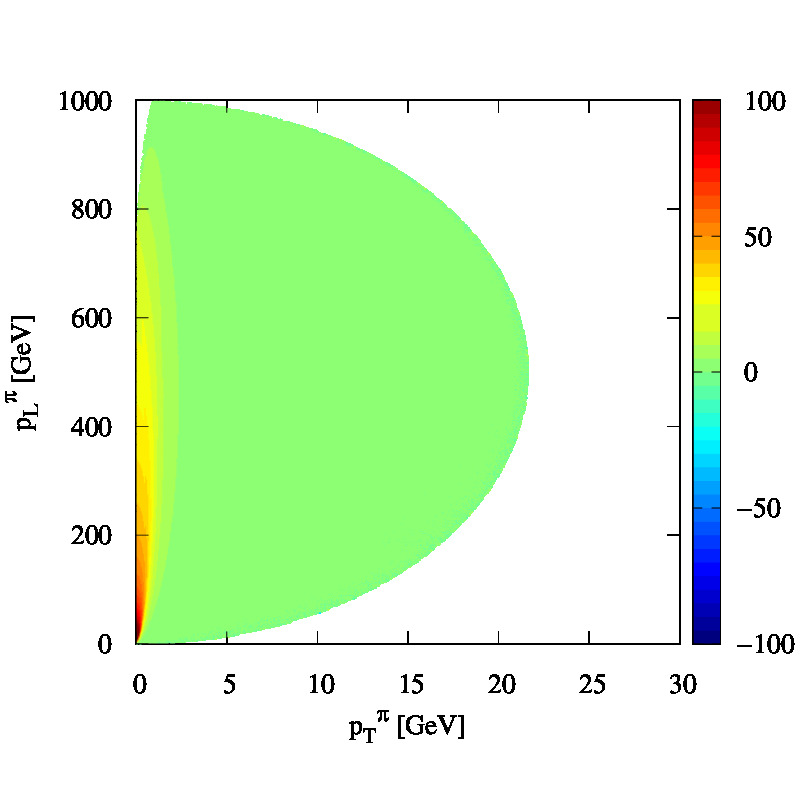} &
    \includegraphics[width=0.5\textwidth]{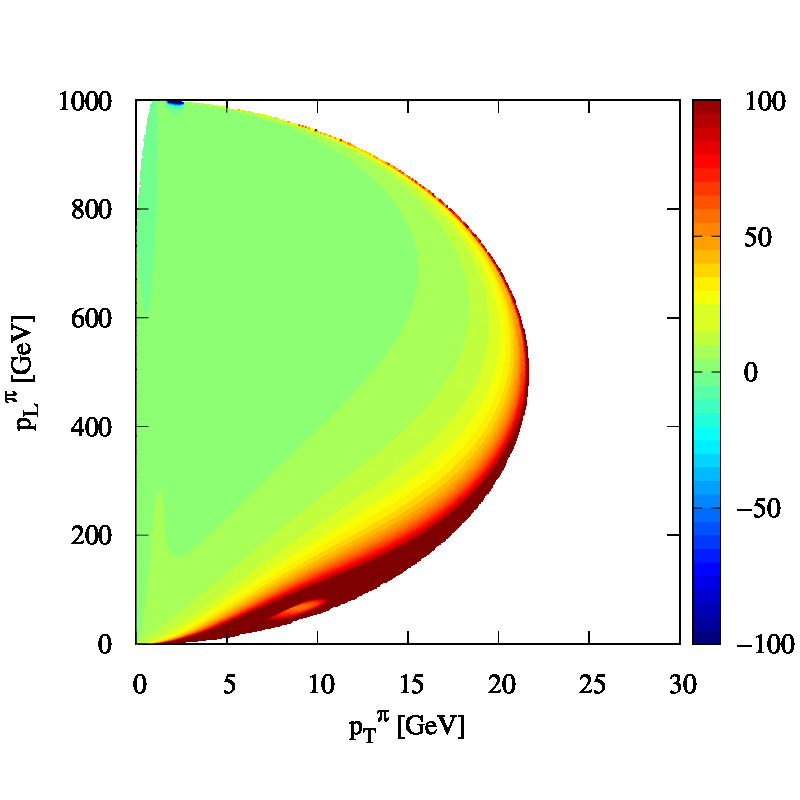}
			\end{tabular}
\caption{ Difference in \% between the predictions for $\mathrm{d}\sigma/(\mathrm{d}p_T^\pi \mathrm{d}p_L^\pi)$ derived assuming the full solution and obtained considering that $F_5 = 0$. Results for an incident (anti)neutrino energy of 100 GeV (upper panels) and 1000 GeV (lower panels). The results are obtained considering the incident (anti)neutrino energy of 100 GeV in the top panels and 1000 GeV in the bottom panels. }
\label{fig:diff_F50}
\end{figure}

In order to estimate the impact of the tau polarization and $F_5$ structure function on this distribution, in what follows we will present our results for the difference in $\%$ between the predictions associated with the full solution with those derived assuming that the (anti) tau is fully polarized (Fig. \ref{fig:diff_PL1}), i.e. $P_L = -1$ for neutrino and $P_L = 1$ for antineutrino, or that $F_5 = 0$ (Fig. \ref{fig:diff_F50}), given by
\begin{eqnarray}
    \begin{aligned}
\frac{\mathrm{d}\sigma/(\mathrm{d}p_T^\pi \mathrm{d}p_L^\pi)|_{F_5=0 (P_L=\pm 1)}-\mathrm{d}\sigma/(\mathrm{d}p_T^\pi \mathrm{d}p_L^\pi)}
{\mathrm{d}\sigma/(\mathrm{d}p_T^\pi \mathrm{d}p_L^\pi)} \times 100\% \, .
    \label{eq:diff_cs}
    \end{aligned}
\end{eqnarray}
From Fig. \ref{fig:diff_PL1} one has that the percentage difference in the differential cross-sections associated with the tau polarization is greater for lower neutrino energies, and therefore for lower tau and pion energies. This behavior is associated with  the fact that tau produced at high energies (much greater than its mass) tends to be a helicity state. Furthermore, we see that the full calculation of tau polarization implies that the distribution is larger than the approximated one for larger transverse momenta and smaller longitudinal momenta.
For processes induced by antineutrinos, again the differences in the distributions are greater for lower energy antineutrinos, but in this case the impact is larger for pions with energy close to zero and close to the energy of the initial antineutrino.

 In Fig. \ref{fig:diff_F50} we present the percentage difference of the double differential cross-sections with the full calculation of Eq. (\ref{eq:sigma2}) and assuming $F_5 = 0$ for an interaction induced by tau (anti)neutrinos. As $F_5$ has a negative sign in the cross-section of Eq. (\ref{eq:sigma2}),  the distribution for an incoming neutrino is decreased in almost the entire pion momentum range when we add this structure function. The inclusion of $F_5$ is important in almost the entire kinematic region allowed for the pion arising from the decay of tau, but its main contribution is on the region of pions with small transverse momentum. Analogously to the case of neutrinos, the contribution of $F_5$ decreases the  distribution associated with an incoming antineutrino in almost the entire momentum range of the pion. On the other hand, the antineutrino case shows  larger percentage differences in the distributions at large transverse moments of the pion when the contribution of  $F_5$ is taken into account. Moreover, one has  that this contribution decreases (increases) the distribution at small (large) values of the  longitudinal momentum.

\section{Summary}
\label{sec:sum}

The expectation of thousands of tau neutrino events at FASER$\nu$2 during the high luminosity era of LHC will allow us to investigate several properties of the third generation of leptons. In this work, motivated also by the fact that the  polarization degree of tau is less than one in GeV-TeV energy regime, we have investigated how this polarization modifies the momentum, energy and angular distributions associated with the pion generated in the (anti) tau decay. We show that the (double) differential cross-section with respect to the pion momentum arising from tau decay is significantly modified if  the tau is assumed fully polarized, with the impact depending on the kinematic regions analyzed. We also investigated the impact of the structure function $F_5$ on the (double) differential cross-sections induced by both neutrinos and antineutrinos, and have verified that the predictions are sensitive to the magnitude of this quantity. Such results indicate that the future experimental analysis of processed induced by (anti) tau neutrinos at the LHC will be useful to improve our understanding of the neutrino properties and the description of the hadronic structure.

\begin{acknowledgments}
The authors thank J. Rojo for a useful comment that has motivated some of the analyses performed in the present study. R. F. acknowledges support from the Conselho Nacional de Desenvolvimento Cient\'{\i}fico e Tecnol\'ogico (CNPq, Brazil), Grant No. 161770/2022-3. V.P.G. was partially supported by CNPq, FAPERGS and INCT-FNA (Process No. 464898/2014-5). D.R.G. was partially supported by CNPq and MCTI.

\end{acknowledgments}

\hspace{1.0cm}

\end{document}